\documentclass[prx, twocolumn,longbibliography]{revtex4-1}

\usepackage{bbm}
\usepackage{amsmath}
\usepackage{hyperref}
\usepackage{graphicx}
\usepackage{graphics}
\usepackage{amssymb}
\usepackage{mathtools}
\usepackage{xcolor}
\usepackage{physics}
\usepackage[british]{babel}
\usepackage{csquotes}
\usepackage{graphicx}
\usepackage{slashed}
\usepackage{mathrsfs}
\usepackage{hyperref}

\hypersetup{
    colorlinks=true,allcolors=blue,}
\graphicspath{{Figs/}}

\usepackage{orcidlink}

\begin{document}
\preprint{APS/123-QED}

\title{Quantum simulation of Fermi-Hubbard model based on transmon qudit interaction}

\author{Arian Vezvaee$^{1}$\orcidlink{0000-0001-7691-2864}}
\email{vezvaee@usc.edu}
\author{Nathan Earnest-Noble$^{2,3}$\orcidlink{0000-0003-1441-9067} }
\author{Khadijeh Najafi$^{2,4}$}
\email{knajafi@ibm.com}

\affiliation{%
$^{1}$Department of Electrical \& Computer  Engineering, and  Center for Quantum Information Science \& Technology,  University of Southern California, Los Angeles, California 90089, USA\\
$^{2}$IBM Quantum, IBM T.J. Watson Research Center, Yorktown Heights, NY 10598 USA\\
$^{3}$IBM Quantum,  Research Triangle Park, NC 27709, USA\\
$^{4}$MIT-IBM Watson AI Lab,  Cambridge MA, 02142, USA
}%


\begin{abstract}

The Fermi-Hubbard model, a fundamental framework for studying strongly correlated phenomena could significantly benefit from quantum simulations when exploring non-trivial settings. However, simulating this problem requires twice as many qubits as the physical sites, in addition to complicated on-chip connectivities and swap gates required to simulate the physical interactions. In this work, we introduce a novel quantum simulation approach utilizing qudits to overcome such complexities. Leveraging on the symmetries of the Fermi-Hubbard model and their intrinsic relation to Clifford algebras, we first demonstrate a Qudit Fermionic Mapping (QFM) that reduces the encoding cost associated with the qubit-based approach. We then describe the unitary evolution of the mapped Hamiltonian by interpreting the resulting Majorana operators in terms of physical single- and two-qudit gates. While the QFM can be used for any quantum hardware with four accessible energy levels, we demonstrate the specific reduction in overhead resulting from utilizing the native Controlled-SUM gate (equivalent to qubit CNOT) for a fixed-frequency ququart transmon. We further transpile the resulting two transmon-qudit gates by demonstrating a qudit operator Schmidt decomposition using the Controlled-SUM gate. Finally, we demonstrate the efficacy of our proposal by numerical simulation of local observables such as the filling factor and Green's function for various Trotter steps. The compatibility of our approach with different qudit platforms paves the path for achieving quantum advantage in simulating non-trivial quantum many-body systems. 
\end{abstract}



\maketitle

\section{Introduction}

\begin{figure*}
\hspace{0cm}{\includegraphics[scale=.65]{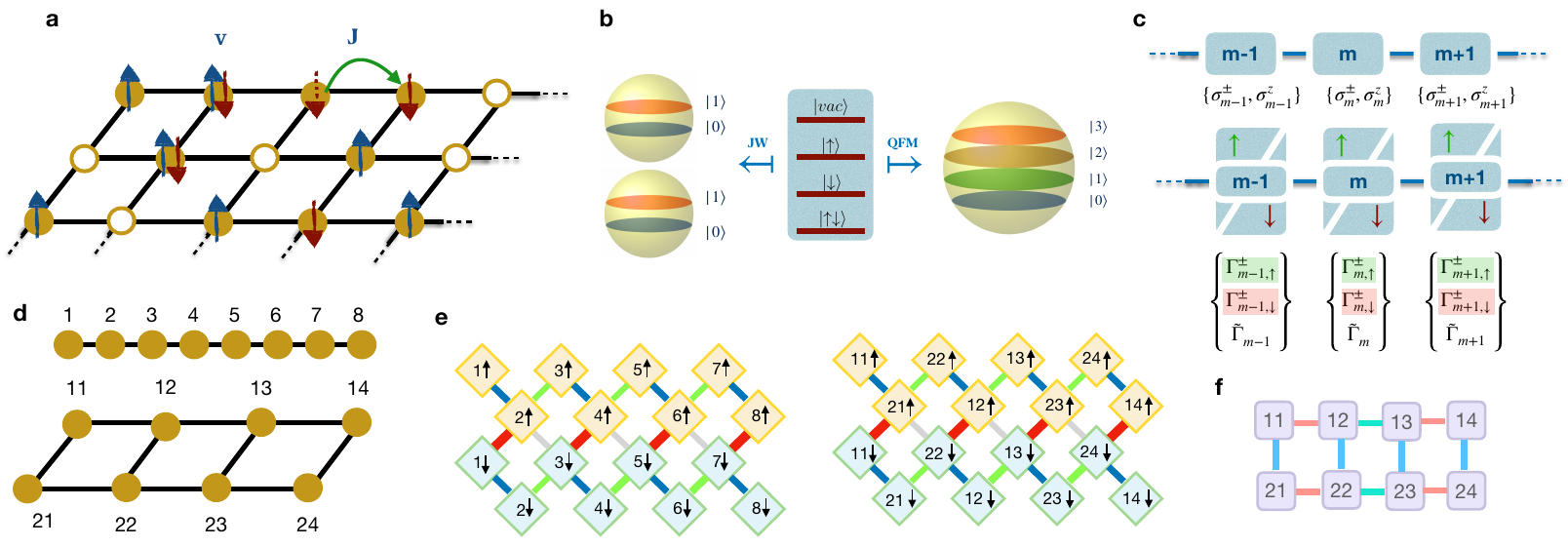}}
\caption{(a) Schematic depiction of a two-dimensional Fermi-Hubbard model with hoping $J$ and on-site interaction $v$. (b) Mapping the four degrees of freedom of a spinful Fermi-Hubbard site onto two two-level systems, versus mapping onto a single ququart. (c) The standard Jordan-Wigner transformation, maps a spinless fermionic chain with two fermionic operators ($a$ and $a^\dagger$), to a combination of spin-1/2 Pauli matrices $\sigma^\pm$. In qudit fermionic mapping, we map the four spinful fermionic operators ($c_\nu$ and $c_\nu^\dagger$ for $\nu=\uparrow,\downarrow$) to the four generators of the Clifford algebra $\mathcal{C}\ell_{0,4}$ (the $\Gamma$ matrices), by associating each spin specie (indicated by green and red) to a subset of two  $\Gamma$ matrices. (d) An example of a $1\times 8$ (top), and a $2\times 4$ (bottom) Fermi-Hubbard lattice. (e) The qubit layout on the zig-zag-like transmon-based chip. The left (right) figure corresponds to the 1$\times$8 ($2\times 4$) lattice. Both cases require sixteen qubits to encode the given Fermi-Hubbard lattices. Colored edges represent the two-qubit gates that can be applied in parallel. In both cases, various SWAP gates are required for performing the hopping and the on-site interaction terms. (f) The qudit fermionic mapping layout for $2\times 4$ Fermi-Hubbard lattice requires only eight qudits. Qudit fermionic mapping preserves the intrinsic connectivity of the lattice Hamiltonian and removes the need for any swap-like operators which significantly reduces the number of required multi-qudit gates.}
\label{fig-schematic}
\end{figure*}

The highly celebrated Fermi-Hubbard model~\cite{Hubbard1963}, a fundamental cornerstone in the study of quantum many-body systems, has been the subject of intense research for several decades~\cite{Arovas2022AnnualReviews}. This model, which describes a system of interacting fermions on a lattice (Fig.~\ref{fig-schematic}(a)), has emerged as a key tool in understanding various strongly correlated phenomena, such as high-temperature superconductivity~\cite{Bednorz1986,Lee2006RevModPhys}, and quantum magnetism~\cite{Anderson1987Science}. In addition, investigation of the rich phase diagram of the two-dimensional Fermi-Hubbard model as a function of doping parameters has led to the development of novel numerical methods~\cite{Gull2015}. Although the Fermi-Hubbard model can be exactly solved in one dimension via the Bethe Ansatz~\cite{Bethe1931,Yang1967PRL}, its solutions in higher dimensions and away from half-filling remain elusive due to the complexity of the many-body wavefunctions and the exponential growth of the Hilbert space. This complexity indeed invites one to explore Feynman's original idea of simulating nature with nature~\cite{Feynman1982} in order to study the properties and behavior of the model in non-trivial settings. In recent years quantum simulation has gained traction with the advances made in quantum hardware~\cite{GeorgescuRevModPhys2014,McArdleRevModPhys2020,Monroe2021RevModPhys2021,Smith2019npj,Kim2023Nature,Dalzell2023arxiv}. In fact, the Fermi-Hubbard model has been simulated both in digital~\cite{Arute2020arxiv,Stanisic2022NatComm} and analog~\cite{Hensgens2017Nature,TARRUELL2018ScienceDirect} setups, where in all these attempts, two-level quantum bits (qubits) have been the primary choice for representing quantum states of the system. While these qubit-based attempts have enabled simulation of the Fermi-Hubbard Hamiltonian, they have often proven to demand a substantial number of qubits and significant quantum operation (gate) with complicated on-chip connectivities to encompass the intrinsic nature of the hoping and interacting terms. Meanwhile, the utilization of higher-dimensional quantum systems, known as qudits~\cite{Lanyon2009NatPhys,Luo2014ScienceChinaPhys,Wang2020FrontierPhys}, has been gaining momentum due to their potential to provide a more efficient representation of quantum states and increased computational power. Several qudit advantages have been demonstrated in various contexts of quantum technologies, namely, quantum error correction with small code size~\cite{Muralidharan2017NJP,Campbell2014PRL}, quantum photonics~\cite{Yoshikawa2018PRA,Zheng2022PRXQuantum,Raissi2022arXiv,Chi2022NatComm}, and quantum information processing~\cite{Naik2017NatComm,Blok2021PRX,Morvan2021PRL}. In particular, due to the fact that qudits provide the ability to
perform simultaneous control on many levels~\cite{Lu2019AdvQuantumaTech}, their multilevel nature has shown great promise in reducing the quantum circuit's complexities. Moreover, qudit advantages in terms of error resilience and information capacity may prove crucial for the development of large-scale quantum simulators~\cite{Neeley2009Science,Tacchino2021JMC,Kurkcuoglu2022arxiv,Cuadra2022PRL,Meth2023arxiv}.

In this paper, we take the first steps to simulate the Fermi-Hubbard model using $d=4$ qudits; i.e., \textit{ququarts}. We leverage the inherent connection of underlying symmetries of the Fermi-Hubbard model with the corresponding Clifford algebra $\mathcal{C}\ell_{0,4}$ to directly map the fermionic degrees of freedom of the Fermi-Hubbard model into quqarts, thus reducing the complexity of the encoding. In particular, since QFM enables mapping each site in the Fermi-Hubbard model to an individual ququart processor (Fig.~\ref{fig-schematic}(b)), our method simplifies the challenges related to the on-chip geometry constraints and the requirement of employing multiple swap gates to execute the Fermi-Hubbard Hamiltonian on a quantum circuit. While QFM is hardware-agnostic, we showcase the implementation of the mapped Hamiltonian on a ququart transmon circuit. We describe the single- and two-qudit gates that correspond to the unitary evolution of the mapped Hamiltonian and provide a recipe for the decomposition of the two-qudit gates in terms of the equivalent CNOT gate for qudits, the Controlled-Sum gate (CSUM). Additionally, many interesting properties of the Fermi-Hubbard model, such as spectral function  and response functions, can be obtained using the Green's function framework. Therefore, we ultimately demonstrate the validity and efficiency of the presented method through numerical simulation of related quantities. 


The paper is structured as follows. In Section~\ref{sec-mapping}, we briefly discuss the physical properties of the Fermi-Hubbard model and present the QFM. In Section~\ref{sec-gates}, we discuss the unitary implementation of the mapped Hamiltonian in terms of single- and two-qudit gates. In Section~\ref{sec-results}, we present numerical simulations of various observables using the mapped Hamiltonian and compare it to the exact results. Finally, we provide a comparison of required resources for qubit versus qudit approaches in Section~\ref{sec-resources}, and conclude in Section~\ref{sec-conclusions}.


\section{Qudit fermionic mapping for the Fermi-Hubbard model} \label{sec-mapping}

We start by laying out some fundamental properties of the Fermi-Hubbard model, and the connection between the spinful fermionic operators of the model and ququarts. While in this section we specifically discuss the mapping between the Fermi-Hubbard model and ququarts, the QFM is generic and is discussed in Appendix~\ref{app-qfm}.

\subsection{The Fermi-Hubbard Hamiltonian} \label{sec-hubbard-properties}

We begin by considering the 1D Fermi-Hubbard Hamiltonian comprising of $L$ number of sites of spinful particles:
\begin{equation} \label{eq-Hubbard-Ham-Original}
    \begin{aligned}
H=  -J \sum_{m=1}^{L-1} \sum_{\nu=\uparrow, \downarrow} (c_{m, \nu}^{\dagger} c_{m+1, \nu}+\text {h.c.}) +\upsilon \sum_{m=1}^L N_{m, \uparrow} N_{m, \downarrow}.
\end{aligned}
\end{equation} Here, the fermionic annihilation (creation) operators for site $m$ and spin degree of freedom $\nu$ is denoted by $c_{m, \nu}~(c_{m, \nu}^{\dagger})$, and $N_{m,\nu} = c_{m, \nu}^{\dagger}c_{m, \nu}$ defines the number operator. The hopping term, characterized by coefficient $J$, represents the \textit{nearest-neighbor} particle tunneling between adjacent sites and the on-site interaction term with coefficient $\upsilon$, takes into account the Coulombic repulsion between charges on the same site (Fig.~\ref{fig-schematic}(a)). The fermionic operators satisfy 
\begin{equation} \label{eq-fermionic-relations}
    \begin{aligned}
& \{c_{m,\nu}, c_{m^\prime,\nu^{\prime}}^{\dagger}\}=\delta_{\nu, \nu^{\prime}} \delta_{m, m^{\prime}}, \\
& \{c_{m,\nu}, c_{m^\prime,\nu^{\prime}}\}=\{c^{\dagger}_{m,\nu}, c^{\dagger}_{m^\prime,\nu^{\prime}}\}=0,
\end{aligned}
\end{equation}
where $\{A,B\}\equiv AB+BA$ defines the anti-commutator. In this picture, only four states per site are possible: $\{\ket{vac}, \ket{\uparrow}, \ket{\downarrow}, \ket{\uparrow\downarrow} \}$, which represent vacuum, a single electron with spin up, a single electron with spin down, and a doubly-occupied up–down pair, respectively. Let us also briefly discuss the symmetry properties of this system. An important property of the Fermi-Hubbard Hamiltonian is the fact that it conserves the spin and charge degrees of freedom. These conservation are associated with the global $\mathrm{SO}(4) \simeq \mathrm{SU}(2) \times \mathrm{SU}(2) / \mathrm{Z}_2$ symmetry of the system~\cite{Yang1991}, which can be extended to local symmetries~\cite{Masumizu2005PRB}; that is, at each site, the $\mathrm{SO}(4)$ transformations can be applied independently. The first $\mathrm{SU}(2)$ (often denoted as the spin symmetry $\mathrm{SU}(2)_{\mathrm{S}}$) is responsible for conservation of the spin operator, and the second $\mathrm{SU}(2)$ (often denoted as the charge symmetry $\mathrm{SU}(2)_{\mathrm{C}}$) reflects the particle-hole symmetry of the system, caused by the electron-hole transformation. As we will see in the following section, this $\mathrm{SO}(4)$ symmetry plays a crucial role in finding an efficient map between the fermionic operators and ququarts: From a pure mathematical point of view, the generators of $\mathrm{SO}(N)$ can be constructed using the operators of the Clifford algebra $\mathcal{C}\ell_{0,N}$~\cite{Georgi1999}. Moreover, as shown in Ref.~\cite{Zhang1991SO4}, interestingly enough, this $\mathrm{SO}(4)$ symmetry is extendable to higher dimensions and it also leads to an exact one-to-one correspondence between the eigenstates at half-filling, and those away from half-filling. This fact allows us to use the QFM for nontrivial settings and multi-dimensional Fermi-Hubbard models.\\

Let us also briefly discuss the physical properties of the Fermi-Hubbard model that are of interest both from theoretical and experimental points of view. Many properties of the Fermi-Hubbard model can be calculated within the Green's function formalism. In particular, the many-body Green's function allows one to evaluate observables such as hopping energy and many many-body densities of states. Below, we briefly explain the Green's function in equilibrium. We start with the Green's function~\cite{Cohn2020PRA}:
\begin{equation} \label{GFR}
\begin{gathered}
G_{i j \nu}^R(t)=-i \theta(t) \frac{1}{\mathcal{Z}} \operatorname{Tr} e^{-\beta H}\left\{c_{i \nu}(t) c_{j \nu}^{\dagger}(0)\right\}, \\
G_{i j \nu}^{<}(t)=i \frac{1}{\mathcal{Z}} \operatorname{Tr} e^{-\beta H} c_{j \nu}^{\dagger}(0) c_{i \nu}(t),
\end{gathered}
\end{equation}
where $\mathcal{Z}=Tr e^{-\beta H}$ is the partition function, and the time-dependent operator is defined in Heisenberg representation $\mathcal{O}(t)=e^{iHt}\mathcal{O}\,e^{-iHt}$. $\theta(t)$ is the Heaviside step function which is zero for $t<0$, 1 for $t>1$, and $0.5$ for $t=0$.  We can also define the Green's function in the frequency domain via the Fourier transform
\begin{equation} \label{eq-GFL}
G_{ij\nu}^{R,<}(\omega)=\int_{-\infty}^{\infty} dt e^{i\omega t}G_{ij\nu}^{R,<}(t).
\end{equation}
It is straightforward to show that the local density of state can be obtained from the imaginary part of the retarded Green's function~\cite{Freericks2019arxiv},
\begin{equation} \label{eq-spectral}
A_{i\nu}(\omega)=-\frac{1}{\pi}\Im{G_{ij\nu}(\omega)}.
\end{equation}


\begin{figure}
\hspace{0cm}{\includegraphics[scale=.28]{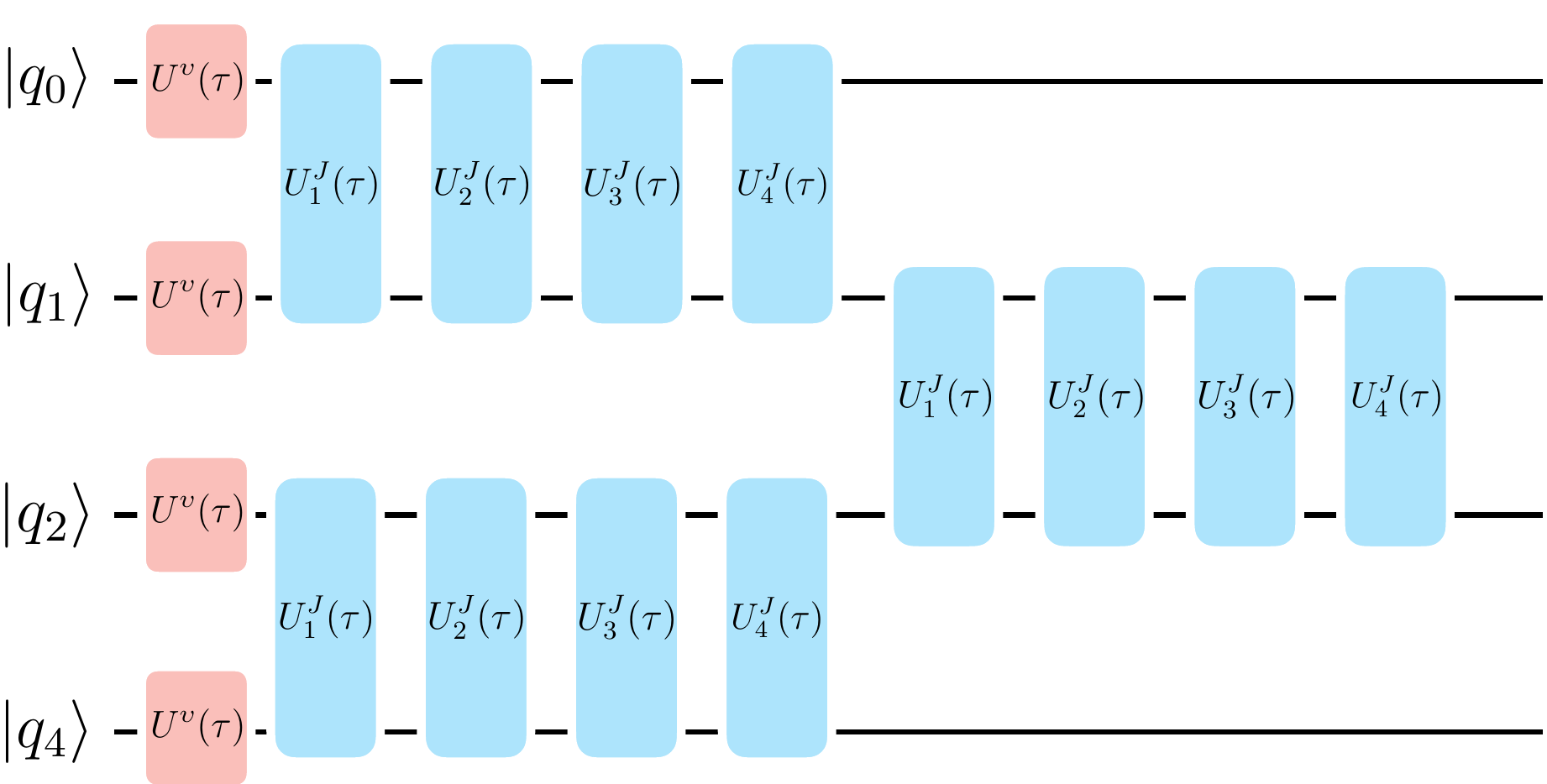}}
\caption{One full Trotter step $\tau$ for a four-site Fermi-Hubbard Hamiltonian mapped onto four ququarts using QFM. Notice that now we only require nearest-neighbor interactions, as in the actual Fermi-Hubbard lattice geometry. Each single- (light pink) and two-qudit gates (light blue) can be expressed in terms of two-level-subspace rotations of the full qudit. }
\label{fig-Trotter-step}
\end{figure}

\subsection{Mapping the Fermi-Hubbard Hamiltonian to qudit hardware}

We begin the discussion of QFM, (that is, mapping the fermionic operators of Fermi-Hubbard Hamiltonian in Eq.~\eqref{eq-Hubbard-Ham-Original} to a ququart), by reviewing the well-known results of Jordan Wigner (JW) transformation for a spinless fermionic chain (Fig.~\ref{fig-schematic}(c)). In this case, the corresponding fermionic relations \eqref{eq-fermionic-relations} are given as, \begin{equation} \label{eq-fermionic-relations-JW}
    \begin{aligned}
& \{a_{m}, a_{m^\prime}^{\dagger}\}= \delta_{m, m^{\prime}}, \\
& \{a_{m}, a_{m^\prime}\}=\{a^\dagger_{m}, a^\dagger_{m^\prime}\}=0.
\end{aligned}
\end{equation}
The JW transformation provides a map that allows one to write these fermionic operators in terms of spin operators while preserving the fermionic commutation relation described above. This (non-local) transformation at site $m$ is given as
\begin{equation}
    a_{m}^{\dagger}=\Pi^{\sigma}_{(m-1)}\sigma_{m}^{+}, \quad a_{m}=\Pi^{\sigma}_{(m-1)} \sigma_{m}^{-},
\end{equation}
where  $\{\sigma_x,\sigma_y,\sigma_z\}$ are the Pauli spin operator matrices at a given site $m$, and $\sigma^\pm=(1/2)(\sigma_x\pm i\sigma_y)$. The \textit{non-local string operator} for site $m$ is defined as $\Pi^{\sigma}_{(m-1)} \equiv \sigma_{1}^{z} \cdots \sigma_{m-1}^{z}$ (with $\Pi^{\sigma}_{0} \equiv \mathbb{I}$). One can apply the JW transformation to the Fermi-Hubbard Hamiltonian for each flavor of the spin individually (Fig.~\ref{fig-schematic}(c)). This results in requiring two qubits per site of the fermionic system: One qubit encodes the spin-up, and one qubit encodes the spin-down. Consequently, simulating the mapped Hamiltonian on quantum hardware will require many two-qubit gates for both the hopping and the on-site interaction terms. Moreover, implementing the hopping term, which involves two sites with different spin degrees of freedom, demands careful engineering of on-chip connections among the qubits and various swap operations to induce interaction among all degrees of freedom~\cite{Arute2020arxiv}. This situation gets even more challenging for two and higher dimensional systems~\cite{Stanisic2022NatComm}.

Here, we devise a QFM that encodes each site of the fermionic system on a single ququart. This may be facilitated by making an important observation: Pauli spin operators are a good choice for the JW mapping as they have the following anti-commutation relation:
\begin{equation} \label{eq-c2}
    \{\sigma_i,\sigma_j\}=2\delta_{i,j}\mathbb{I}_{2\times 2}.
\end{equation}
The key concept here is that the Pauli spin operators $\sigma_{x(y)}$ share similar algebraic properties to the operators of the $\mathcal{C}\ell_{0,2}$ Clifford algebra. An $N$-dimensional Clifford algebra associated with a Euclidean space, $\mathcal{C}\ell_{0,N}$, is defined by a set of operators satisfying, \begin{equation}\label{eq-C4-algebra}
   \{\Gamma_i,\Gamma_j\}=+2 \delta_{i,j}, \enskip \text{for} \enskip i, j=1,\cdots,N.
\end{equation}Therefore the JW transformation can be seen as mapping the fermionic operators of a spinless chain with $U(1)$ symmetry (equivalent to $SO(2)$) to the Clifford operators of $\mathcal{C}\ell_{0,2}$. 
With this notion in mind, one can find a QFM by noting that the fermionic anti-commutation relations of spinful chains, and the anti-commutation relations of the corresponding Clifford algebra (Eq.~\eqref{eq-C4-algebra}) can be mapped to one another~\cite{Masumizu2005PRB}. Thus, on each site $m$ of the Fermi-Hubbard model, we consider a set of fermionic operators ($c_{m, \nu},~c_{m, \nu}^{\dagger}$) that are mapped to a set of local Clifford operators associated with site $m$~(Fig.~\ref{fig-schematic}(c)). In this QFM, one can loosely think of the local Clifford operators at each site $\Gamma_i^m$, to play the role of the local Pauli spin operators. 

As we discussed in Section~\ref{sec-hubbard-properties}, in the case of the Fermi-Hubbard model with two spin degrees of freedom, the symmetry group is $\mathrm{SO}(4)$ with the corresponding Clifford algebra of $\mathcal{C}\ell_{0,4}$. This algebra provides us with four generating operators denoted as $\Gamma_i$, for $i=1,2,3,4$, where all these operators mutually anti-commute according to Eq.~\eqref{eq-C4-algebra}. Additionally, there exists a fifth element of this algebra $\Tilde{\Gamma}$ which anti-commutes with the set above and is the result of the product of the four $\Gamma$ matrices: \begin{equation}\Tilde{\Gamma}\equiv-\Gamma_1\Gamma_2\Gamma_3\Gamma_4.\end{equation} This $\Tilde{\Gamma}$ matrix can be seen as the counterpart of the $\sigma_z$ spin operator that constructs the string term in JW. 
We now divide the four $\Gamma$ matrices into two subsets of two, and use each subset to map a single spin degree of freedom: \begin{eqnarray} \label{eq-GJW}
c_{m,\uparrow}^\dagger \;\mapsto\; \frac{1}{2}\,(\tilde\Gamma^1 \cdots \tilde\Gamma^{m-1})(\Gamma_1^m + i \Gamma_2^m), \nonumber \\
c_{m,\uparrow} \;\mapsto\; \frac{1}{2}\,(\tilde\Gamma^1 \cdots \tilde\Gamma^{m-1})(\Gamma_1^m - i \Gamma_2^m),  \\
c_{m,\downarrow}^\dagger \;\mapsto\; \frac{1}{2}\,(\tilde\Gamma^1 \cdots \tilde\Gamma^{m-1})(\Gamma_3^m + i \Gamma_4^m), \nonumber \\
c_{m,\downarrow} \;\mapsto\; \frac{1}{2}\,(\tilde\Gamma^1 \cdots \tilde\Gamma^{m-1}) (\Gamma_3^m - i \Gamma_4^m). \nonumber
\end{eqnarray}

It is straightforward to show that this mapping satisfies the fermionic anti-commuting relations in Eq.~\eqref{eq-fermionic-relations}. Under this map, the hopping and the on-site interaction terms of the Fermi-Hubbard Hamiltonian in Eq.~\eqref{eq-Hubbard-Ham-Original} transform as $\tilde{H}=\tilde{H}_\text{Hop} +\tilde{H}_\text{Int}$, where (see Appendix~\ref{app-qfm} for the derivations of each term),
\begin{equation} \label{eq-hopping-mapped}
  \tilde{H}_\text{Hop} = \frac{J}{2i}\sum_{m=1}^{L-1}\sum_{\ell=2,4}\!  \enskip({\Gamma}_\ell^m\tilde{\Gamma}^m \otimes\Gamma_{\ell-1}^{m+1}-{\Gamma}_{\ell-1}^m\tilde{\Gamma}^m \otimes\Gamma_{\ell}^{m+1}),
\end{equation} and,
\begin{equation} \label{eq-on-site-mapped}
 \tilde{H}_\text{Int} =  4\upsilon\sum_{m=1}^{L}(\mathbb{I} -i \Gamma_1^m\Gamma_2^m-i \Gamma_3^m\Gamma_4^m + \tilde{\Gamma}^m).
\end{equation}
We should note the following crucial observations. Under this map, the summation over the spin degree of freedom in Eq.~\eqref{eq-Hubbard-Ham-Original} for the hopping term, has now transformed into a summation over the four $\Gamma$ matrices in Eq.~\eqref{eq-hopping-mapped}; that is, we have mapped a spinful fermionic Hamiltonian into a spinless chain. Furthermore, it can be seen immediately that under this transformation, the on-site interaction term remains local within each ququart. We show in Section~\ref{sec-gates} that this leads to gate operations on a single qudit as opposed to the qubit-based approach where many two-qubit and swap-like gates are required to simulate the on-site term (see e.g., Ref.~\cite{Arute2020arxiv}). Moreover, as we discuss in the following section, in the case of ququart transmons, the required single qudit gates to implement this term can be done trivially with unit fidelity. Additionally, we should note that while the hopping term requires two-qudit interactions, it only requires a nearest-neighbor interaction as in the actual Fermi-Hubbard model, thus reducing any complexity associated with the geometry design and the on-chip couplings of the quantum processors. In other words, QFM allows us to perform a digital quantum simulation of the Fermi-Hubbard model while precisely mimicking the intrinsic connectivity of the original Hamiltonian (Fig.~\ref{fig-schematic}(d)).

Thus far we have utilized the $\Gamma$ matrices only in their purely algebraic form in order to write down our mapping. To find a proper representation for the $\Gamma$ matrices, we should recall that the Clifford algebra that we use here is associated with an Euclidean space. Therefore, bases such as Dirac or Weyl do not lead to the correct fermionic algebra and give rise to anti-Hermitian operators due to their construction~\footnote{Note that in this case Eq.~\eqref{eq-C4-algebra} reads as $ \{\Gamma_i,\Gamma_j\}=2 \eta_{i,j}$ where $\eta_{i,j}$ is the metric tensor, associated with the time- and space-like components of the Clifford algebra $\mathcal{C}\ell_{p,q}$}. Consequently, we need to choose an appropriate representation basis for the $\Gamma$ matrices such that it assures not only the fermionic relations hold correctly but also facilitates implementation of the mapped Hamiltonian on our quantum hardware. One such choice of basis is indeed the Majorana basis given below:

\begin{eqnarray} \label{eq-Gamma-basis}
    \Gamma_1&=&\sigma_x\otimes\mathbb{I}_{2\times 2}, \quad
    \Gamma_2=\sigma_y\otimes\mathbb{I}_{2\times 2}, \nonumber \\
    \Gamma_3&=&\sigma_z\otimes\sigma_x, \quad \quad
    \Gamma_4=\sigma_z\otimes\sigma_y, \\
    \tilde\Gamma&=&\sigma_z\otimes\sigma_z=-\Gamma_1\Gamma_2\Gamma_3\Gamma_4. \nonumber
\end{eqnarray}
Each $\Gamma$ matrix above can be seen as an operation on a ququart. We will discuss how each of these Majorana operators can be implemented on our qudit hardware in the following section.

\begin{figure*}
\hspace{0cm}{\includegraphics[scale=.27]{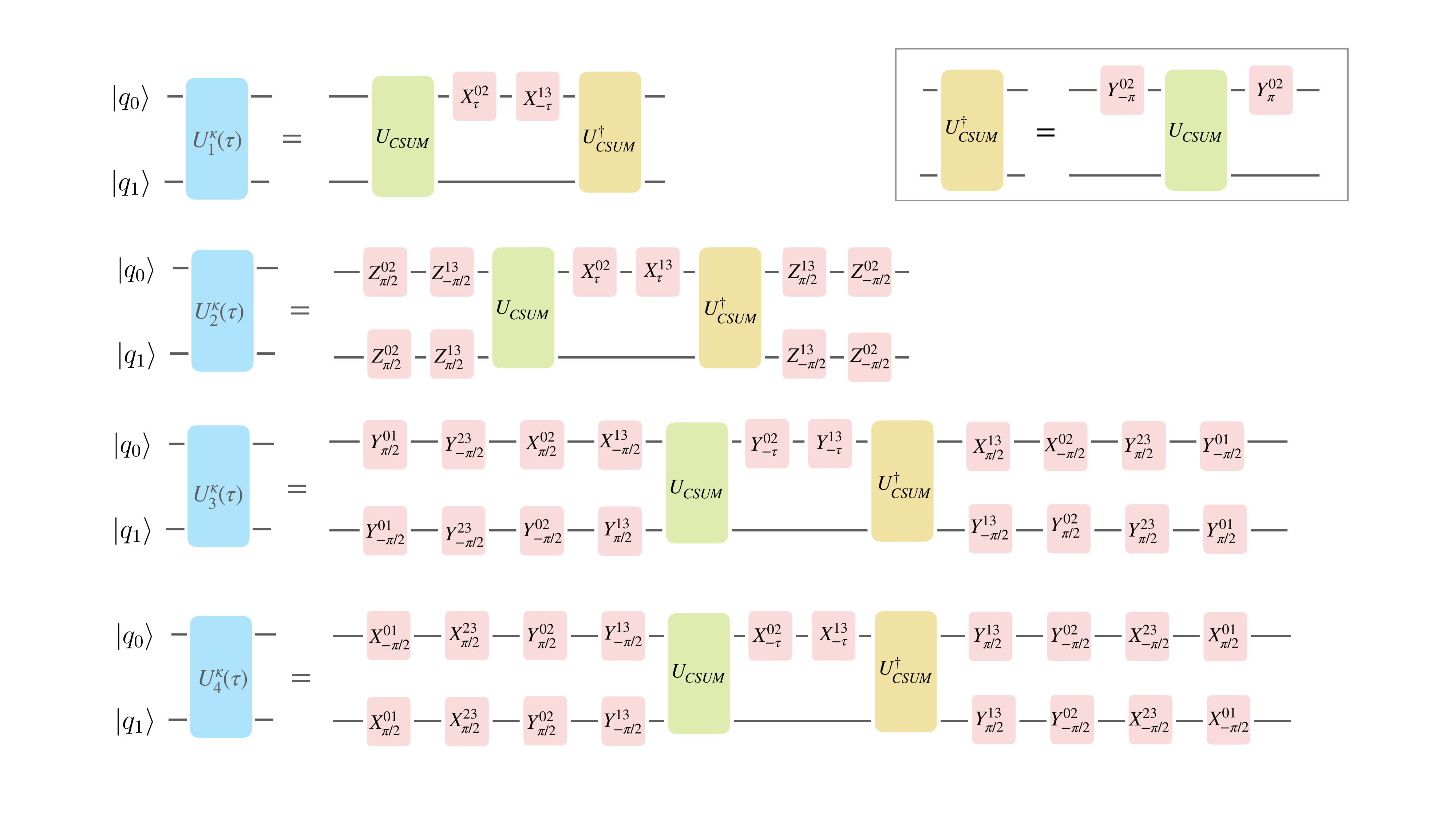}}
\caption{Decomposition of the two-qudit gates given in a single step of Trotterization (Fig.~\ref{fig-Trotter-step}) in terms of many single qudit gates (light pink) and the $U_{\rm{CSUM}}$ gate. The $U_{\rm{CSUM}}$ gate is the equivalent of a CNOT gate, extended to the qudit systems. The top right inset shows the decomposition of the two-qudit gate $U^\dagger_{\rm{CSUM}}$ in terms of the native interaction of the hardware, the $U_{\rm{CSUM}}$ gate, and available single-qudit gates.  }
\label{fig-two-qudit}
\end{figure*}



\section{Single- and two-qudit gates for the mapped Hamiltonian} \label{sec-gates}

The unitary implementation of the mapped Hamiltonian and corresponding gate decomposition is considered the key step in quantum simulations. Here, we proceed with interpreting the resulting terms in Eqs.~\eqref{eq-hopping-mapped} and \eqref{eq-on-site-mapped} in terms of single- and two-qudit gates. We first discuss the Trotterized Hamiltonian that needs to be implemented on the quantum circuit, and then we present the generic recipe for performing rotations in a single multi-level qudit and the relevance of $\Gamma$ matrices to this scheme. We should emphasize that all of our results up to this point are hardware agnostic and can be implemented in a variety of qudit-based processors. Only in Section~\ref{sec-csum} where we discuss the two-qudit gates and their native interactions, do we make hardware-specific assumptions about utilizing transmon ququarts for our gate decomposition and simulations. We ultimately discuss the potential quantum control methods that allow for the implementation of the quantum gates that we obtain in Section~\ref{sec-gate-implementation}, 


\subsection{Trotterization} \label{sec-Trotterization}

The mapped Hamiltonian $\tilde{H}=\tilde{H}_\text{Hop}+\tilde{H}_\text{Int}$ under QFM is comprised of two terms that do not commute: $[\tilde{H}_\text{Hop},\tilde{H}_\text{Int}]\neq 0$. Therefore we have to Trotterize our Hamiltonian as $e^{-i\hat{H} \tau}\approx\left(e^{-i \tilde{H}_\text{Hop} \tau / n} e^{-i \tilde{H}_\text{Int} \tau / n}\right)^n.$ One Trotter step for a four-site example is shown in Fig.~\ref{fig-Trotter-step}. As mentioned in the previous section, the $\tilde{H}_\text{Int}$ term is a local interaction that corresponds to a series of single-qudit operations. The $\tilde{H}_\text{Hop}$ term in Eq.~\eqref{eq-hopping-mapped}, on the other hand, corresponds to interactions among adjacent qudits, which necessarily requires the implementation of two-qudit gates. In the following, we will describe the implementation of both the single- and two-qudit gates and their decompositions.

\subsection{Single qudit rotations in a multi-level system}

In a generic $d$-level system, rotations between any two levels can be accomplished by embedding the Pauli matrices into two-dimensional subspaces of the $d$-dimensional structure. Specifically, we can define
\begin{eqnarray}    
x^{j k}&\equiv&|j\rangle\langle k|+| k\rangle\langle j|, \nonumber\\
y^{j k}&\equiv& -i|j\rangle\langle k|+i| k\rangle\langle j|, \\
z^{j k}&\equiv&|j\rangle\langle j|-| k\rangle\langle k|, \nonumber
\end{eqnarray}
for $0 \leq j<k\leq d$. From this set of operators, the $x^{j k}$ and $y^{j k}$ correspond to transitions between levels $\ket{j}$ and $\ket{k}$, thus we can perform Rabi oscillations in the corresponding subspace. Similarly, the $z^{j k}$ corresponds to the virtual Z gates~\cite{McKayPRA2017} within the two-level subspace of the qudit. For a $d=3$ qutrit, the operators above are often called the Gell-Mann operators~\cite{Blok2021PRX,Morvan2021PRL}. We label the $d>3$ case as the Generalized Gell-Mann (GGM) operators. We also reserve the lowercase $\{x,y,z\}$ to denote these operators for a given subspace of the $d=4$ ququart, and denote the corresponding \textit{rotation} with an angle $\phi$ as, \begin{equation} \label{eq-GGM-rotations}
    X^{jk}_\phi \equiv e^{-i\frac{\phi}{2} x^{jk}  }, \enskip  Y^{jk}_\phi \equiv e^{-i\frac{\phi}{2} y^{jk}  }, \enskip  Z^{jk}_\phi \equiv e^{-i\frac{\phi}{2} z^{jk} }. 
\end{equation} Moreover, we should note that the commutation relation among each two levels also generalizes to this multi-level picture (e.g., $[x^{jk},y^{\ell p}]=2iz^{jk}\delta_{j\ell}\delta_{kp}$). This allows us to safely perform single-qudit gates among every two sublevels of the system in any desired order. We can now describe each of the $\Gamma$ matrices in our mapped Hamiltonian, in terms of the qudit operators that we have defined above:
\begin{eqnarray} \label{eq-gamma-composition}
\Gamma_1&=&x^{20}+x^{31},~\Gamma_2=y^{02}+y^{13},\nonumber \\
\Gamma_3&=&x^{01}-x^{32},~  \Gamma_4=y^{01}+y^{32}, \\
\tilde{\Gamma}&=&z^{01}-z^{23}.\nonumber
\end{eqnarray} The evolution of the mapped Hamiltonian can now be understood in terms of the GGM rotations (i.e.,~Eqs.~\eqref{eq-GGM-rotations}) among two-level subspaces of the full ququart. We should also note that any combination of two $\Gamma$ matrices will result in a new operator that can be written in terms of GGM operators. As such, for the mapped on-site interaction term in Eq.~\eqref{eq-on-site-mapped}, we can write, \begin{equation} \label{eq-on-site-gates}
 \tilde{H}_\text{Int} \sim  4\upsilon(\mathbb{I}+z^{01}+z^{02}+z^{03}),
\end{equation} which implies that in a single step of Trotterizaton (Fig.~\ref{fig-Trotter-step}), the evolution that corresponds to the on-site interaction term will be
\begin{equation} U^\upsilon(\tau)=Z^{01}_{2\upsilon\tau}Z^{02}_{2\upsilon\tau}Z^{03}_{2\upsilon\tau},
\end{equation}
up to the global phase, arising from the constant term. In the following section, we will describe the mapped hopping term with regards to the composition of $\Gamma$ matrices in Eq.~\eqref{eq-gamma-composition}, which, unlike the on-site interaction term, will require two-qudit gates among the nearest-neighbor ququarts.


\subsection{Two-qudit interactions and gate decomposition} \label{sec-csum}

As we previously mentioned, the form of the mapped qudit Hamiltonian is generic and independent of the choice of hardware. Moreover, the single qudit rotations that we discussed in the preceding section are also universal and can be implemented in any desired platform. The two-qudit interactions, however, are quantum processor-specific, as the nature of two-qudit interaction and the means to achieve them varies in different platforms. We should note that while the results we present here are specific to transmon qudits, the underlying methodology of translating the mapped Hamiltonian in Eq.~\eqref{eq-hopping-mapped} into two-qudit gates remains the same for other platforms. This methodology is as follows: In any given platform, first, we need to identify the source of our two-qudit interactions. This source of interaction will act as the \textit{native gate} that generates entanglement between any two qudits in the system. Once the native gate is set, we can proceed to decompose the two-qudit terms in the mapped hopping term of the Fermi-Hubbard model, in terms of the native gate and additional single-qudit gates. We will show in the following that this procedure can be done systematically using the Operator-Schmidt Decomposition (OSD)~\cite{Tyson2003IOP1,Tyson2003IOP2} for qudits.

In what follows we describe the procedure above for transmon qudits. An attractive aspect of working with transmons is that the native interaction can be designed for more optimal choices of gate decompositions, unlike other systems such as molecular qudits where the interactions are typically fixed by the nature of the quantum system~\cite{Najafi2019JCPL,Tacchino2021JMC}. Consequently, different methods for achieving two-qudit interaction for qudit transmons are available, which could depend on the geometry and the coupling scheme of adjacent superconducting qudits. In particular, if the neighboring qudit transmons are capacitively coupled to one another, they are subject to an always-on \textit{cross-Kerr} interaction, which we can harness to perform our two-qudit operations. It is shown in Ref.~\cite{Blok2021PRX} that one can combine single qudit operations and the built-in cross-Kerr interaction to perform a Controlled-SUM (CSUM) gate between two adjacent $d$-dimensional qudits:\begin{equation}  \label{eq-UCSUM}
    U_{\text{CSUM}}=\sum_{n=1}^d |n\rangle\langle n| \otimes \tilde{X}^n,
\end{equation}
where $\tilde{X}$ is the qudit operator,
\begin{equation}
\tilde{X}=\sum_{j=1}^d |j\rangle\langle j+1|,
\end{equation}such that $\tilde{X}^d=\mathbb{I}_{d\times d}$. As such, we can think of the $U_{\rm{CSUM}}$ gate as the qudit analog of the qubit Controlled-NOT gate: Depending on which state the control qudit is in, the second qudit will be subject to $\tilde{X},~\tilde{X}^2,\cdots,\tilde{X}^d=\mathbb{I}_{d\times d}$. We choose this $U_{\text{CSUM}}$ gate as our native gate, due to its similar universality~\cite{Su2022PRA} to the CNOT gate. That is, while we are relying on the fact that this gate has been implemented experimentally on transmons, due to its universality, it is expected to be achieved in other qudit-based systems as well. (Although, to the best of our knowledge, we are not aware of any experimental implementation of $U_{\text{CSUM}}$ in other platforms). In principle, however, one could benefit from specific choices of interactions between the qudits to simplify the decomposition procedure (for more detail see the discussion in Appendix~\ref{app-two-qudit-decomposition}).

The procedure to perform OSD for two-qudit gates using the $U_{\text{CSUM}}$ is as follows. Given a target arbitrary two-qudit unitary evolution $U_{\mathbb{C}1,\mathbb{C}2}$, we perform OSD to decompose the given unitary $U_{\mathbb{C}1,\mathbb{C}2}=\sum_{i=0}^n \lambda_i A_i \otimes B_i$, where $\lambda$'s are the corresponding Schmidt coefficients. Next, we need to perform OSD on an ansatz unitary that involves $U_{\text{CSUM}}$ and compare the resulting decomposition with the target decomposition. Depending on how different the two decompositions are, we can add additional single-qudit operators until we reach the target decomposition. In our case, we start by identifying the terms from Eq.~\eqref{eq-hopping-mapped} as $h_1^m\equiv -i{\Gamma}_2^m\tilde{\Gamma}^m \otimes\Gamma_{1}^{m+1}$, $h_2^m\equiv i{\Gamma}_1^m\tilde{\Gamma}^m \otimes\Gamma_{2}^{m+1}$, $h_3^m\equiv -i{\Gamma}_4^m\tilde{\Gamma}^m \otimes\Gamma_{3}^{m+1}$, and $h_4^m\equiv i{\Gamma}_3^m\tilde{\Gamma}^m \otimes\Gamma_{4}^{m+1}$. Thus we define our target evolutions $U_i^J(\tau)=e^{-i h_iJ \tau}$. We then apply OSD to each term to analyze the target two-qudit evolution. With this target evolution as a reference, we can perform OSD on an ansatz that contains only the $U_{\text{CSUM}}$ gate and single qudit gates. If required, we then incorporate additional single-qudit gates to achieve the desired target evolution. We show the results of this decomposition in Fig.~\ref{fig-two-qudit}. We leave the details of the OSD procedure to Appendix~\ref{app-two-qudit-decomposition}.


\begin{figure}
\hspace{0cm}{\includegraphics[scale=.85]{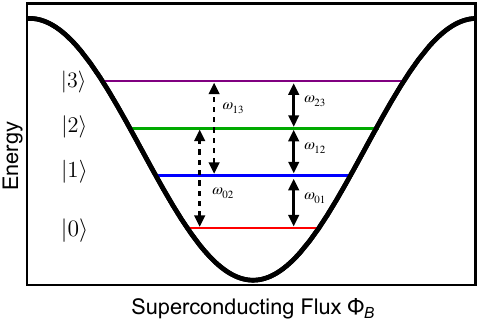}}
\caption{Schematic depiction of the energy levels of a superconducting transmon ququart and the required types of single-qudit gates for a single step of Trotterization. The solid arrows represent a gate that involves two adjacent energy levels of a ququart and the dashed arrows represent single-qudit gates that involve two non-adjacent levels.    }
\label{fig-transmon-gates}
\end{figure}
\subsection{Gate implementation in trasmon ququarts} \label{sec-gate-implementation}

Superconducting circuit-based transmons facilitate a favorable platform for quantum information processing, owing to their high gate and measurement repetition rates, and relative longevity in coherence times~\cite{Bravyi2022JAP}. Additionally, their unique control mechanism, based on an anharmonic oscillator (Fig.~\ref{fig-transmon-gates}), offers a valuable tool for executing both qubit and qudit gates~\cite{Galda2021arxiv,Fischer2022PRResearch,Blok2021PRX,Seifert2023arxiv,Fischer2023PRX}. This anharmonic oscillator is achieved through coupling a Josephson junction, with energy $E_J$, to a capacitor with energy $E_C$. The anharmonic multilevel structure of the transmons creates a natural environment for encoding qudits. While transmons higher levels are prone to charge noises, they can be engineered such that $E_J/E_C \gg 1$. This approach dramatically mitigates charge noise that could otherwise trigger undesirable fluctuations in the eigenenergies of the system, all while maintaining adequate anharmonicities to allow for individual transition control~\cite{Blais2021RMP}. Current IBM transmon devices have relatively low $E_J/E_C$ ratios in the range of 40-50 which induces charge noise fluctuations of around 20 MHz in the $\ket{2}\leftrightarrow\ket{3}$ transition~\cite{Fischer2023PRX}, while other groups have developed transmon devices with a $E_J/E_C \approx 80$ that retain high fidelity qubit-based gates but have a dramatically improved charge dispersion~\cite{goss2022high}. Here we consider ideal transmon devices which are not affected by charge noise, which is supported as a reasonable assumption from recent experiments that have shown to be robust against charge noises for both qutrits~\cite{Blok2021PRX}, and ququarts~\cite{Liu2023PRX} when $E_J/E_C \approx 80$. Additionally, various recent studies have been done for achieving high-fidelity control schemes in ququarts~\cite{Fischer2022PRResearch,Fischer2023PRX,Liu2023PRX}. 

The required gates for a single step of Trotterization are a combination of the two-qudit CSUM, and a series of single-qudit gates that involve either two adjacent energy levels of a transmon, or two non-adjacent levels (Fig.~\ref{fig-transmon-gates}). The rotations that involve the adjacent energy levels can be implemented by driving the corresponding transition in a standard manner. Moreover, the $Z$ rotations among arbitrary energy levels can be implemented with unit fidelity through \textit{virtual $Z$} gates~\cite{McKayPRA2017}, which are shown to be generalizable to qudits~\cite{Fischer2022PRResearch,Liu2023PRX}. However, since one cannot coherently drive all transitions of transmon qudit using a single pulse, we have to implement the non-adjacent rotations in a different manner. The non-adjacent rotations can be achieved by sequentially combining two perpendicular axes of rotations (i.e., $X$- and $Y$-rotations) among adjacent energy levels. Consequently, $X$-rotation among two non-adjacent levels $X_\phi^{m,m+2}$ can be implemented as 
\begin{equation}
    X_\phi^{m,m+2} = Y_{-\pi}^{m,m+1}X_\phi^{m+1,m+2}Y_{\pi}^{m,m+1},
\end{equation}and similarly, for the $Y$-rotation among two non-adjacent levels \begin{equation}
    Y_\phi^{m,m+2} = X_{-\pi}^{m,m+1}Y_\phi^{m+1,m+2}X_{\pi}^{m,m+1}.
\end{equation} 
Each individual adjacent-level rotation can be done within the Derivative Removal by Adiabatic Gates (DRAG)~\cite{MotzoiPRL2009, GambettaPRA2011,Vezvaee2023PRX} formalism to improve the fidelity of each gate by mitigating the effects of the leakage to the levels not involved in the operation. Moreover, we should note that $X$- and $Y$-rotations are also related to one another through a virtual Z operation~\cite{McKayPRA2017}. Therefore, in our simulations, we assume that we have access to only the $X$ gates, and perform the $Y$-rotations as $X$ gates accompanied by the proper virtual Z gates.


\section{Numerical simulations for two-point Green's Functions} \label{sec-results}

To showcase the success of our method, in this section we present the results from emulating the qudit circuit for the two-, three-, and four-site Fermi-Hubbard model using transmon ququarts. We compare the results to those obtained by performing exact numerical calculations for various properties of the Fermi-Hubbard model, such as the occupation number, as well as the lesser Green's function.

\begin{figure*}
\hspace{0cm}{\includegraphics[scale=.7]{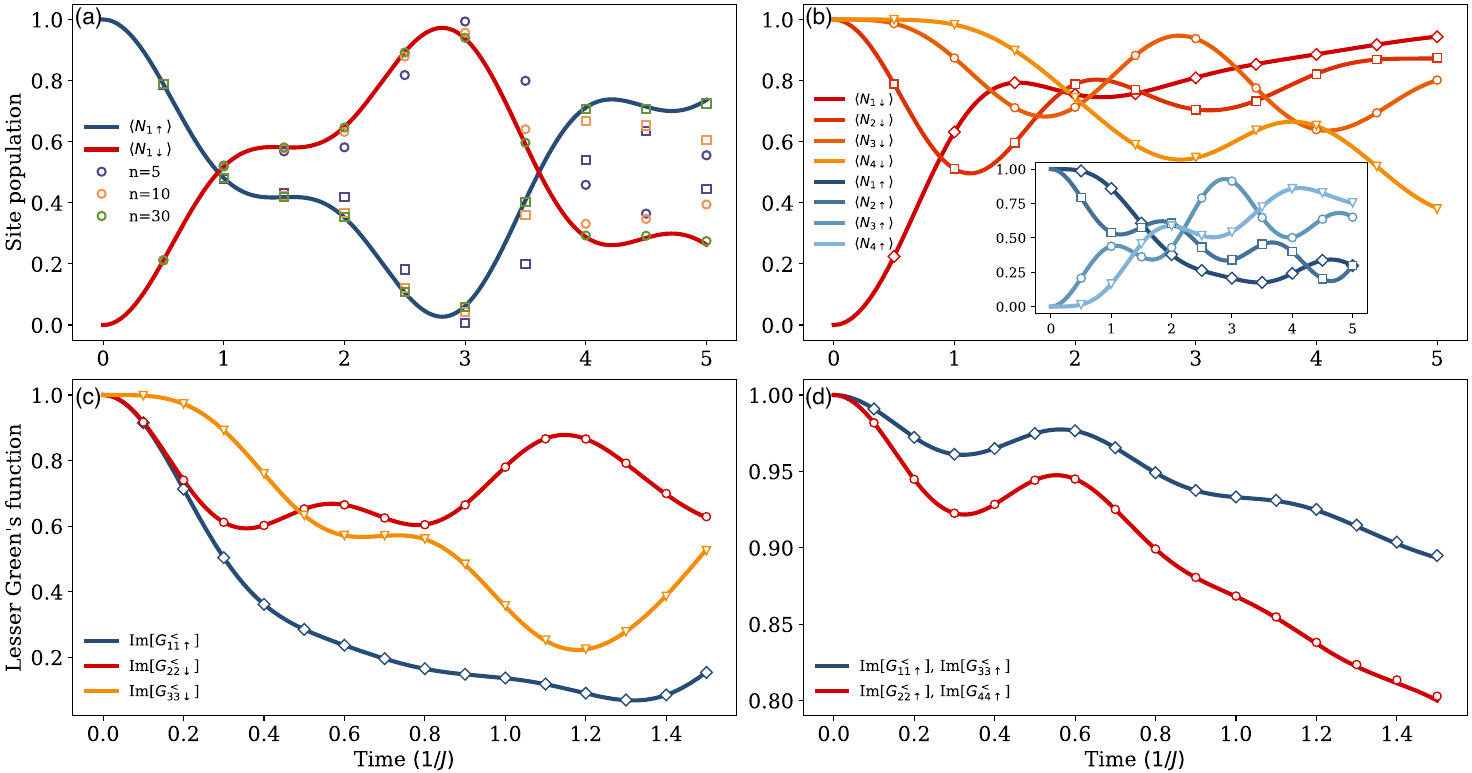}}
\caption{Simulation results for the population dynamics of (a) two-site Fermi-Hubbard model at half filling and (b) a four-site example with initial state $\ket{\psi(t=0)}=\ket{\uparrow,\uparrow\downarrow,\uparrow,\downarrow}$. In each example, the continuous curves represent the exact diagonalization results, and the open markers represent the specific values that are obtained by emulating the qudit circuit ($0.5\leq \tau\leq 5.0$ with intervals of $\delta \tau=0.5$). In (a) green, purple, and orange represent Trotterization steps 5, 10, and 30, respectively. In (b) only the Trotterizatoin step of 30 is shown. In (c) we show exact (solid lines) and emulation results (open markers) of the LGF for a four-site Fermi-Hubbard model, and in (d) we show similar results for the Greens function.}
\label{fig-res-1}
\end{figure*}

Figures~\ref{fig-res-1}(a) and (b) show a simple study of the population dynamics of each site, i.e., $\langle N_{m,\nu}(t)\rangle=\langle c_{m,\nu}^\dagger(t) c_{m,\nu}(t) \rangle$, with respect to the two spin degrees of freedom for the two- and four-site Fermi-Hubbard model. To benchmark the result of our quantum simulations, we first compute the population dynamics of the system using an exact numerical method (via QuTiP~\cite{Johansson2012QuTiP}). The open markers correspond to the Trotterized circuit for a given time $t=\tau/J$, with various Trotterization step numbers $n$, indicated with different markers. In both cases, we have simulated $0.5\leq \tau/J\leq 5.0$ with intervals of $\delta \tau=0.5$. In Fig.~\ref{fig-res-1}(a), we have simulated a two-site Fermi-Hubbard system at half-filling. That is, the initial state of the system corresponds to $\psi(t=0)=\ket{\uparrow,\downarrow}$. Due to the conservation of spin and charge (see Section~\ref{sec-hubbard-properties}) in this case, the two sites will simply exchange the two particles over time. Notice that in the figure we have only shown the population at the first site, $\langle N_{1,\uparrow}\rangle$ and $\langle N_{1,\downarrow}\rangle$. The population of the second site, $\langle N_{2,\uparrow}\rangle$ and $\langle N_{2,\downarrow}\rangle$, will follow the exact same trajectory over time. The markers in green, purple, and orange represent Trotterization steps $n=5, 10, 30$, respectively. As seen from the figure, at $n=30$ we are able to simulate the population dynamics efficiently. In Fig.~\ref{fig-res-1}(b), we have simulated a four-site Fermi-Hubbard system with the initial state of $\ket{\psi(t=0)}=\ket{\uparrow,\uparrow\downarrow,\uparrow,\downarrow}$ which departs from the half filling condition and the population dynamics demonstrate a non-trivial behavior. Once again, the Trotterization step size of $n=30$ (shown with various open markers for each site) is capable of efficiently capturing the correct dynamics of the system. Finally, we show the quantum simulation of the lesser Green's Function (Eq.~\eqref{eq-GFL}) in Fig.~\ref{fig-res-1}(c) and (d). In Fig.~\ref{fig-res-1}(c) we only show the three non-zero lesser Green's Functions computed for a chain of three ququarts, and similarly, we show the two non-zero lesser Green's Function for the four sites in In Fig.~\ref{fig-res-1}(d). In both cases, we are using Trotterization step size $n=30$ which once again efficiently captures the correct numerically-calculated dynamics.


\section{Resource estimation and comparison with qubit mapping} \label{sec-resources}

In this section, we consider the computational resource estimates and the potential gain from the simulation of the Fermi-Hubbard model using qudits compared to the traditional qubit methods. We start by noting that based on the two-qudit decompositions shown in Fig.~\ref{fig-two-qudit}, and keeping in mind that the on-site interaction term only requires virtual Z gates, each Trotterization step for two adjacent sites requires exactly 32 \textit{non-virtual-Z} single qudit operations. Considering that every single operation takes roughly $\sim 50$ ns on average (different transitions require different gate times to avoid leakage and other errors), given the most recent decoherence limits of ququarts ~\cite{Liu2023PRX,Seifert2023arxiv}, Trotterization step size of $n\sim 10$ should be conveniently achievable on current devices. However, we hope that the presented work stimulates interest in further study and improvement of suitable quantum hardware.

Next, we consider the comparison of the required number of two-qudit gates versus two-qubit gates for the two approaches. To formally quantify this comparison we consider the two experiments done using transmon qubits for two Fermi-Hubbard systems of size 1$\times$8 and 2$\times$4 using the ``zig-zag" configuration (Fig.~\ref{fig-schematic}(e)) in Ref.~\cite{Stanisic2022NatComm}. Firstly, while in both cases we would need sixteen qubits, using QFM only eight qudits are required, which as shown in Fig.~\ref{fig-schematic}(f), preserves the physical layout of the Fermi-Hubbard lattice under study. This gives an overall advantage in terms of connectivity and device geometry, which could be specifically crucial for a superconducting-based experiment. Next, we should note that to evolve the system under the full Fermi-Hubbard Hamiltonian using this zig-zag setup, one needs to perform two-qubit Fermionic SWAP (FSWAP) gates~\cite{KivlichanPRL2018} to prevent the implementation of long-distance four-qubit operations. The presence of various layers of FSWAP gates inevitably increases the number of required two-qubit gates for a single Trotterization step. In particular, for the 1$\times$8 experiment, we require the following order of two-qubit operations: FSWAP (red), on-site terms (blue), FSWAP (red), odd hopping terms (blue), even hopping terms (green). Given the gate decomposition in Ref.~\cite{Stanisic2022NatComm}, this brings the total number of two-qubit gates to 64. Compared to the decomposition we used in Fig.~\ref{fig-two-qudit}, we would only need 56 two-qudit gates for the same experiment. We should also note that this is the decomposition based on the $U_{\text{CSUM}}$ native interaction; the two-qudit gate count can be improved by taking advantage of engineering the interaction between the qudits to closely resemble the mapped-Hamiltonian nearest-neighbor interaction. This could potentially help to implement the full mapped hopping term using less number of two-qudit gates compared to a system that only has access to $U_{\text{CSUM}}$ native interaction (see Appendix~\ref{app-two-qudit-decomposition}). Nevertheless, the qudit-based approach will prove to be superior compared to the qubit-based approach in the $2\times 4$ case. The overhead of two-qubit gates can be obtained from the order of operations for evolving the system under the full Hamiltonian: FSWAP (red), on-site terms (blue), FSWAP (red), vertical hopping terms (blue), FSWAP (blue), first set of horizontal hopping terms (green), FSWAP (blue), second set of horizontal hopping terms (green). This requires a total number of 112 two-qubit gates. The QFM on the other hand, once again, renders all the on-site terms local and only needs 10 hopping terms shown in Fig.~\ref{fig-schematic}(f) which would require a total of 80 two-qudit gates. This would make an enormous difference once we consider the size of Fermi-Hubbard lattices that are hard to solve classically and can benefit from quantum simulations.


\section{Conclusions and discussions} \label{sec-conclusions}

In this work, we demonstrated the efficacy of qudit-based quantum simulations by studying the Fermi-Hubbard model and comparing our results with those obtained from exact numerical methods. The QFM described in our work presents a general framework to sufficiently simulate the Fermi-Hubbard model using $d=4$ ququarts, which paves the way for exploring its applicability on various quantum platforms. The QFM approach, compared to the corresponding qubit-based approach, not only reduces the encoding cost but also significantly reduces the complexities associated with the on-chip connectivity and geometry of the quantum device. That is, due to the fact that QFM renders each site as a single quantum processor, it only requires nearest-neighbor connectivity. This is a critical requirement for developing robust schemes for extending the quantum simulations to higher dimensions as well as non-trivial settings that lead to actual quantum advantage. That is, while the Fermi-Hubbard model is considered a top candidate for near-term demonstration of quantum advantage due to being less demanding compared to quantum simulation of, e.g. molecules~\cite{Dalzell2023arxiv}, our results highlight the potential versatility of using qudits for reducing the experimental overhead even further to gain valuable insights into the underlying physics of strongly correlated systems. Moreover, we have demonstrated the extension of the OSD method to the qudit case and presented the decomposition of the resulting mapped Hamiltonian in terms of universal qudit gates. We particularly considered the $U_{\text{CSUM}}$ gate as the native interaction between the adjacent qudits since it corresponds to the universal CNOT gate in the qubit setting. This avoids the limitation of our presented work to only superconducting circuit-based transmons and facilitates the implementation of this approach on other quantum platforms, such as trapped ions~\cite{Monroe2021RevModPhys2021,Blatt2012NaturePhysics,nam2019groundstate}, Rydberg atoms~\cite{Gonz_lez_Cuadra_2023}, and single magnetic molecules~\cite{Baldovi2015IOP,Najafi2019JCPL,Chiesa2021AIP,Tacchino2021JMC}. While in this work we have only focused on the dynamical properties, our method can be utilized to investigate the ground state properties for the Fermi-Hubbard model, given the fact that variational quantum eigensolvers (VQEs) leverage on Hamiltonian to build efficient variational anstaze~\cite{Wecker2015,Cade2020,Stanisic2022NatComm}. 
This would be similar to Ref.~\cite{Stanisic2022NatComm} where efficient variational ansatz and the Trotterized Fermi-Hubbard Hamiltonian are utilized to observe properties such as metal-insulator transition and Friedel oscillations of system size of eight, in one and two dimensions. 

Our work establishes a new paradigm for simulating spinful quantum many-body systems by leveraging the higher dimensionality of qudit systems. Moreover, due to the high degrees of freedom, qudits are ideal for describing gauge fields which are naturally higher dimensional. While recent works have focused on investigating various gauge fields in qudits~\cite{meth2023simulating,Zache_2023,Gonzlez_Cuadra_2022,halimeh2022stabilizing}, the presented method, in particular, can pave the path toward gauge fields that interact with fermionic matter. Given the recent surge in exploring qudit-based systems for various purposes, we are optimistic that the advantages of using qudits for quantum simulations presented in our work, further stimulate the interest in developing the necessary tools for utilizing qudits for quantum information processing. In particular, future work could focus on the design of robust single- and two-qudit gates, as well as exploring the qudit setting for the fault-tolerant quantum simulation of the Fermi-Hubbard model~\cite{Campbell2021IOP}.

%


\section*{Acknowledgements} 
The authors express their deepest gratitude to Sergey Bravyi for his valuable insights, and for suggesting the term `Qudit Fermionic Mapping'. The authors would like to thank Ivano Tavernelli, Peter Zoller, Hannes Pichler, Martin Ringbauer, Jad C. Halimeh, Francesco Tacchino, Torsten V. Zache, Ehsan Khatami, Semeon Valgushev, Charles Headfield, Nathan Leitao, Edwin Barnes, and Sophia Economou for helpful discussions and constructive feedback at various stages of completing this work. 

\appendix

\section{Qudit Fermionic Mapping} \label{app-qfm}
In this appendix, we present the QFM in its general form, where the QFM transformation takes a form similar to the usual JW transformation: \begin{equation}
    \begin{gathered}
c_{m,\nu}^\dagger \;\mapsto\; \frac{1}{2}\,(\tilde\Gamma^{1} \cdots \tilde\Gamma^{m-1}) \Gamma_{m,\nu}^{+}, \\
c_{m,\nu} \;\mapsto\; \frac{1}{2}\,(\tilde\Gamma^{1} \cdots \tilde\Gamma^{m-1}) \Gamma_{m,\nu}^{-}, 
\end{gathered}
\end{equation}here, we have defined $
\Gamma_{m,\nu}^{\pm} \equiv \frac{1}{2}\left(\Gamma^m_{2 \nu-1} \pm i \Gamma^m_{2 \nu}\right).
$ For the Fermi-Hubbard case, we can only have $\nu=1,2$ since we have two spin species, i.e., we have taken the convention that $\nu: 1\mapsto\uparrow,\quad
2\mapsto\downarrow$. In this case, the formal QFM transformation turns into the results in Eq.~\eqref{eq-GJW}. Notice that the map above removes the spin degree of freedom and essentially allows for the simulation of a spinless model. Let us consider a generic hopping term with spin up, $\nu=1$. Under this mapping, 
\begin{eqnarray}
c_{m,\uparrow}^\dagger c_{m+1,\uparrow} &=&[\tilde\Gamma^1...\tilde\Gamma^{m-1}] \Gamma_{m,1}^{+} [\tilde\Gamma^1...\tilde\Gamma^{m-1}\tilde\Gamma^{m}]  \Gamma_{m+1,1}^{-} \nonumber  \\
&=&  [(\tilde\Gamma^1)^2...(\tilde\Gamma^{m-1})^2][\Gamma_{m,1}^{+} \tilde\Gamma^m][\Gamma_{m+1,1}^{-}]\nonumber \\
&=&[(\Gamma_1^m + i \Gamma_2^m) \tilde\Gamma^m] [\Gamma_1^{m+1} - i \Gamma_2^{m+1}],
\end{eqnarray}where the direct product of different sites $m$ is implictly implied. Therefore it is straightforward to show that a similar calculation holds for the spin down $\nu=2$ and conclude that the generic form for the hopping is given as in Eq.~\eqref{eq-hopping-mapped}. Similarly, we can work out the interaction term using the QFM transformation as,\begin{eqnarray} \label{eq-on-site}
  N_{m,\uparrow} N_{m,\downarrow} &=& c_{m, \uparrow}^{\dagger} c_{m, \uparrow} c_{m, \downarrow}^{\dagger} c_{m, \downarrow} \\
  &=& [\bigotimes_{k=1}^{m-1} (\tilde\Gamma^k)^4]\otimes[\Gamma_{m,1}^+\Gamma_{m,1}^-\Gamma_{m,2}^+\Gamma_{m,2}^- ] \nonumber \\
   &=&4(1 -i \Gamma_1^m\Gamma_2^m-i \Gamma_3^m\Gamma_4^m + \tilde\Gamma^m), \nonumber
\end{eqnarray}which up to the constant factor corresponds to Eq.~\eqref{eq-on-site-mapped}.

\section{Two-qudit gate decompositions} \label{app-two-qudit-decomposition}
In this section, we present the OSD method for transpiling a two-qudit gate into a series of single qudit gates and the native interaction between the two qudits. We will showcase the method by presenting how one can see the results from Fig.~\ref{fig-two-qudit}. We should note that while we specifically work with the $U_{\text{CSUM}}$ gate defined in Eq.~\eqref{eq-UCSUM} for ququart transmons, the OSD method is generic and can be applied for other types of native interaction among qudits.

We demonstrate the methodology described in Section~\ref{sec-csum}  by showcasing how to find the gates shown in Fig.~\ref{fig-two-qudit}. For concreteness, we provide the OSD results for each of the four target hopping terms: 
\begin{widetext}
\begin{eqnarray} \label{eq-osd-term-1}
U_1^j(\tau)&=&\sin(\tau)\left(
\begin{array}{cccc}
 0 & 0 & -i & 0 \\
 0 & 0 & 0 & i \\
 -i & 0 & 0 & 0 \\
 0 & i & 0 & 0 \\
\end{array}
\right) \otimes \left(
\begin{array}{cccc}
 0 & 0 & 1 & 0 \\
 0 & 0 & 0 & 1 \\
 1 & 0 & 0 & 0 \\
 0 & 1 & 0 & 0 \\
\end{array}
\right) + \cos(\tau) \left(
\begin{array}{cccc}
 1 & 0 & 0 & 0 \\
 0 & 1 & 0 & 0 \\
 0 & 0 & 1 & 0 \\
 0 & 0 & 0 & 1 \\
\end{array}
\right) \otimes \left(
\begin{array}{cccc}
 1 & 0 & 0 & 0 \\
 0 & 1 & 0 & 0 \\
 0 & 0 & 1 & 0 \\
 0 & 0 & 0 & 1 \\
\end{array}
\right), \\ \label{eq-osd-term-2}
U_2^j(\tau)&=&\sin(\tau)\left(
\begin{array}{cccc}
 0 & 0 & -i & 0 \\
 0 & 0 & 0 & i \\
 i & 0 & 0 & 0 \\
 0 & -i & 0 & 0 \\
\end{array}
\right) \otimes \left(
\begin{array}{cccc}
 0 & 0 & -1 & 0 \\
 0 & 0 & 0 & -1 \\
 1 & 0 & 0 & 0 \\
 0 & 1 & 0 & 0 \\
\end{array}
\right) + \cos(\tau) \left(
\begin{array}{cccc}
 1 & 0 & 0 & 0 \\
 0 & 1 & 0 & 0 \\
 0 & 0 & 1 & 0 \\
 0 & 0 & 0 & 1 \\
\end{array}
\right) \otimes \left(
\begin{array}{cccc}
 1 & 0 & 0 & 0 \\
 0 & 1 & 0 & 0 \\
 0 & 0 & 1 & 0 \\
 0 & 0 & 0 & 1 \\
\end{array}
\right), \\ \label{eq-osd-term-3}
U_3^j(\tau)&=&\sin(\tau)\left(
\begin{array}{cccc}
 0 & i & 0 & 0 \\
 i & 0 & 0 & 0 \\
 0 & 0 & 0 & i \\
 0 & 0 & i & 0 \\
\end{array}
\right) \otimes \left(
\begin{array}{cccc}
 0 & -1 & 0 & 0 \\
 -1 & 0 & 0 & 0 \\
 0 & 0 & 0 & 1 \\
 0 & 0 & 1 & 0 \\
\end{array}
\right) + \cos(\tau) \left(
\begin{array}{cccc}
 1 & 0 & 0 & 0 \\
 0 & 1 & 0 & 0 \\
 0 & 0 & 1 & 0 \\
 0 & 0 & 0 & 1 \\
\end{array}
\right) \otimes \left(
\begin{array}{cccc}
 1 & 0 & 0 & 0 \\
 0 & 1 & 0 & 0 \\
 0 & 0 & 1 & 0 \\
 0 & 0 & 0 & 1 \\
\end{array}
\right), \\ \label{eq-osd-term-4}
U_4^j(\tau)&=&\sin(\tau)\left(
\begin{array}{cccc}
 0 & i & 0 & 0 \\
 -i & 0 & 0 & 0 \\
 0 & 0 & 0 & i \\
 0 & 0 & -i & 0 \\
\end{array}
\right) \otimes \left(
\begin{array}{cccc}
 0 & 1 & 0 & 0 \\
 -1 & 0 & 0 & 0 \\
 0 & 0 & 0 & -1 \\
 0 & 0 & 1 & 0 \\
\end{array}
\right) + \cos(\tau) \left(
\begin{array}{cccc}
 1 & 0 & 0 & 0 \\
 0 & 1 & 0 & 0 \\
 0 & 0 & 1 & 0 \\
 0 & 0 & 0 & 1 \\
\end{array}
\right) \otimes \left(
\begin{array}{cccc}
 1 & 0 & 0 & 0 \\
 0 & 1 & 0 & 0 \\
 0 & 0 & 1 & 0 \\
 0 & 0 & 0 & 1 \\
\end{array}
\right). 
\end{eqnarray} \end{widetext}Here, we have written everything in the full Hilbert space of the two qudits as $\mathcal{H}_{Q1}\otimes \mathcal{H}_{Q2}$. We should also note that the OSD second term for all the decompositions above is $\mathcal{I}(\tau)\equiv \cos(\tau) I \otimes I$. This signals that our anstaze (up to additional single qudit gates) essentially only needs to produce the correct first term in each case. For instance if we use the ansatz, $U_{\rm{CSUM}}(X_\tau^{02}X_{-\tau}^{13}\otimes I)U_{\rm{CSUM}}^\dagger$, we retrieve the exact same evolution as Eq.~\eqref{eq-osd-term-1}. Therefore no additional gates are required. We follow the same procedure for the $U_2^j(\tau)$, and we find that performing OSD on the ansatz $U_{\rm{CSUM}}(X_\tau^{02}X_\tau^{13}\otimes I)U_{\rm{CSUM}}^\dagger$, compared to Eq.~\eqref{eq-osd-term-2}, we find,
\begin{equation}
    \sin(\tau)\left(
\begin{array}{cccc}
 0 & 0 & -i & 0 \\
 0 & 0 & 0 & i \\
 i & 0 & 0 & 0 \\
 0 & -i & 0 & 0 \\
\end{array}
\right) \otimes \left(
\begin{array}{cccc}
 0 & 0 & -1 & 0 \\
 0 & 0 & 0 & -1 \\
 1 & 0 & 0 & 0 \\
 0 & 1 & 0 & 0 \\
\end{array}
\right)+\mathcal{I}(\tau).
\end{equation}It is straightforward to see that using the combination of single-qudit gates given in Fig.~\ref{fig-two-qudit}, we retrieve the full evolution from Eq.~\eqref{eq-osd-term-2}. We follow the exact same strategy for the remaining two terms. For completeness, we present the OSD results for the anstaze used for $U_3^j(\tau)$ and $U_4^j(\tau)$. That is for $U_{\rm{CSUM}}(Y_{-\tau}^{02}Y_{-\tau}^{13}\otimes I)U_{\rm{CSUM}}^\dagger,$ and $U_{\rm{CSUM}}(X_{-\tau}^{02}X_{-\tau}^{13}\otimes I)U_{\rm{CSUM}}^\dagger$, we respectively have,
\begin{eqnarray}
\sin(\tau)\left(
\begin{array}{cccc}
 0 & 0 & 1 & 0 \\
 0 & 0 & 0 & 1 \\
 -1 & 0 & 0 & 0 \\
 0 & -1 & 0 & 0 \\
\end{array}
\right) \otimes \left(
\begin{array}{cccc}
 0 & 0 &1 & 0 \\
 0 & 0 & 0 & 1 \\
 1 & 0 & 0 & 0 \\
 0 & 1 & 0 & 0 \\
\end{array}
\right)+\mathcal{I}(\tau),\nonumber\\ \\
\sin(\tau)\left(
\begin{array}{cccc}
 0 & 0 & i & 0 \\
 0 & 0 & 0 & i \\
 i & 0 & 0 & 0 \\
 0 & i & 0 & 0 \\
\end{array}
\right) \otimes \left(
\begin{array}{cccc}
 0 & 0 & 1 & 0 \\
 0 & 0 & 0 & 1 \\
 1 & 0 & 0 & 0 \\
 0 & 1 & 0 & 0 \\
\end{array}
\right)+\mathcal{I}(\tau). \nonumber \\
\end{eqnarray}We should re-emphasize that while the OSDs of the four hopping terms are unique, the solutions given in Fig.~\ref{fig-two-qudit} are not. Indeed, one can use a different ansatz to reduce the length of required single-qudit gates. More importantly, here we are using $U_{\rm{CSUM}}$ as the native two-qudit interaction due to its universality. Ideally, one could engineer the interaction between the two qudits to resemble the target Hamiltonian hopping term to reduce the required two-qudit gates by combining multiple terms of $h_1^m,...,h_4^m$ into a single term and decompose the resulting evolution as a single term in the Trotterized Hamiltonian. This, however, requires studying the potential achievable interactions in the specific hardware used for the simulation. Moreover, while this approach might reduce the number of two-qudit gates (which are potentially the most error-prone operations on the circuit) it might as well require a longer chain of single-qudit gates to achieve the combined evolution resulting from the combination of $h_1^m,...,h_4^m$. As such, there might be a trade-off between the number/fidelity of two-qudit gates, and the length of one Torotterization step compared to the decoherence limit of qudits.


\newpage
\bibliography{biblo}

\end{document}